%% file: content/00-main.tex
\documentclass[sigconf,screen]{acmart}

\usepackage{enumitem}
\usepackage{amsmath,amsfonts}
\usepackage{algorithmic}
\usepackage{array}
\usepackage[caption=false,font=normalsize,labelfont=sf,textfont=sf]{subfig}
\usepackage{textcomp}
\usepackage{stfloats}
\usepackage{url}
\usepackage{verbatim}
\usepackage{graphicx}
\usepackage{balance}
\usepackage{adjustbox}
\usepackage{booktabs}
\usepackage{multirow}
\usepackage{threeparttable}
\usepackage{arydshln}
\usepackage{tcolorbox}
\usepackage{soul}
\usepackage{fontawesome}
\usepackage{makecell}
\usepackage{natbib}
\usepackage{colortbl}
\usepackage{siunitx}
\usepackage{microtype}
\definecolor{mygray}{gray}{.9}
\usepackage{tikz}
\usepackage[T1]{fontenc}
\usepackage{aecompl}

\AtBeginDocument{
  \providecommand\BibTeX{{
    \normalfont B\kern-0.5em{\scshape i\kern-0.25em b}\kern-0.8em\TeX}}}

\newcommand{\Comment}[1]{}

\begin{document}

\title{A Preliminary Study of Large Language Models for Multilingual Vulnerability Detection}

\author{Junji Yu}
\orcid{0009-0001-1147-3685}
\affiliation{
  \institution{College of Intelligence and \\
  Computing, Tianjin University}
  \city{Tianjin}
  \country{China}
}
\email{junjiyu@tju.edu.cn}

\author{Honglin Shu}
\orcid{0009-0005-7311-7060}
\affiliation{
  \institution{Kyushu University}
  \city{Fukuoka}
  \country{Japan}
}
\email{shu.honglin.167@s.kyushu-u.ac.jp}

\author{Michael Fu}
\orcid{0000-0001-7211-3491}
\affiliation{
  \institution{The University of Melbourne}
  \city{Melbourne}
  \country{Australia}
}
\email{michael.fu@unimelb.edu.au}

\author{Dong Wang}
\authornotemark[1]
\orcid{0000-0002-2004-0902}
\affiliation{
  \institution{College of Intelligence and \\
  Computing, Tianjin University}
  \city{Tianjin}
  \country{China}
}
\email{dong_w@tju.edu.cn}

\author{Chakkrit Tantithamthavorn}
\orcid{0000-0002-5516-9984}
\affiliation{
  \institution{Information Technology, Monash University}
  \city{Clayton}
  \country{Australia}
}
\email{chakkrit@monash.edu}

\author{Yasutaka Kamei}
\orcid{0000-0002-7058-1045}
\affiliation{
  \institution{Kyushu University}
  \city{Fukuoka}
  \country{Japan}
}
\email{kamei@ait.kyushu-u.ac.jp}

\author{Junjie Chen}
\orcid{0000-0003-3056-9962}
\affiliation{
  \institution{College of Intelligence and \\
  Computing, Tianjin University}
  \city{Tianjin}
  \country{China}
}
\email{junjiechen@tju.edu.cn}

\thanks{*Dong Wang is the corresponding author.}

\begin{abstract}
Deep learning-based approaches, particularly those leveraging pre-trained language models (PLMs), have shown promise in automated software vulnerability detection.
However, existing methods are predominantly limited to specific programming languages, restricting their applicability in multilingual settings.
Recent advancements in large language models (LLMs) offer language-agnostic capabilities and enhanced semantic understanding, presenting a potential solution to this limitation.
While existing studies have explored LLMs for vulnerability detection, their detection performance remains unknown for multilingual vulnerabilities.
To address this gap, we conducted a preliminary study to evaluate the effectiveness of PLMs and state-of-the-art LLMs across seven popular programming languages.
Our findings reveal that the PLM CodeT5P achieves the best performance in multilingual vulnerability detection, particularly in identifying the most critical vulnerabilities.
Based on these results, we further discuss the potential of LLMs in advancing real-world multilingual vulnerability detection.
This work represents an initial step toward exploring PLMs and LLMs for cross-language vulnerability detection, offering key insights for future research and practical deployment.
\end{abstract}



\acmYear{2025}\copyrightyear{2025}
\setcopyright{acmlicensed}
\acmConference[ISSTA Companion '25]{34th ACM SIGSOFT International Symposium on Software Testing and Analysis}{June 25--28, 2025}{Trondheim, Norway}
\acmBooktitle{34th ACM SIGSOFT International Symposium on Software Testing and Analysis (ISSTA Companion '25), June 25--28, 2025, Trondheim, Norway}
\acmDOI{10.1145/3713081.3731746}
\acmISBN{979-8-4007-1474-0/25/06}

\maketitle

\input{content/01-introduction}
\input{content/02-background}
\input{content/03-evaluation}
\input{content/04-result}

\input{content/05-discussion}
\input{content/06-threat}

\section{Conclusion}
\label{sec:conclusion}
In this paper, we present a preliminary study of PLMs and LLMs for multilingual vulnerability detection.
Our experimental results demonstrate that CodeT5P achieves the best overall performance among the evaluated models.
Through in-depth analysis of model performance across different programming languages, we observe that PLMs maintain superior performance on most languages, while LLMs exhibit particularly strong capabilities in C/C++ vulnerability detection.
Regarding the top 25 most dangerous CWE-IDs in 2024, CodeT5P shows the best detection accuracy, with closed-source LLMs outperforming their open-source counterparts.
These findings underscore two key insights: 
(1) PLMs currently demonstrate robust capabilities for multilingual vulnerability detection, and (2) LLMs show promising potential in this domain, particularly for certain languages and vulnerability types.
However, utilizing universal LLMs for vulnerability detection presents challenges due to the requirement for extensive expert knowledge.
Future research should focus on developing advanced prompt engineering strategies and integrating corporate vulnerability knowledge to improve the detection accuracy of LLMs in vulnerability detection task.


\begin{acks}
This work was supported by National Key Research and Development Program of China (Grant No. 2024YFB4506300), the National Natural Science Foundation of China (Grant No. 62322208), JST under the Adopting Sustainable Partnerships for Innovative Research Ecosystem (ASPIRE) program, Grant Number JPMJAP2415, JSPS for the KAKENHI grants (JP21H04877, JP22K18630), Bilateral Program grant JPJSBP120239929, and the Inamori Research Institute for Science for supporting Yasutaka Kamei via the InaRIS Fellowship.
\end{acks}

\balance
\begingroup
\Large
\bibliographystyle{ACM-Reference-Format}
\bibliography{reference/reference}
\endgroup

\end{document}

%% file: content/01-introduction.tex
\section{Introduction}
Software vulnerabilities refer to coding weaknesses that may be exploited to breach system security, escalate privileges, or trigger malicious outcomes.
For example, the vulnerability CVE-2024-5565 affects a library that generates SQL queries using LLMs with retrieval-augmented generation (RAG).
This library employs a prompt function to present users with visualized query results. However, it is susceptible to prompt injection attacks, allowing an adversary to manipulate the prompt and execute arbitrary Python code instead of the intended visualization code.
With the rise of artificial intelligence, deep learning-based methods~\cite{dam2018automatic, yuan2023enhancing, wu2022vulcnn} have been applied to the field of vulnerability detection.  
Automated Vulnerability Detection (AVD) methods~\cite{fu2022linevul} construct datasets related to vulnerability detection and train models to learn the characteristic patterns of vulnerabilities, thereby predicting whether a given code contains vulnerabilities.  
Compared to traditional methods (i.e., static analysis~\cite{10.1145/3576915.3624401} and dynamic analysis~\cite{kim2019rvfuzzer}), they demonstrate significant advantages in generality.  

With the increasing complexity of modern software systems, multi-language hybrid development has become a standard practice.
Many projects, directly or indirectly, rely on diverse software packages (i.e., software ecosystems), further contributing to this trend.
However, existing deep learning-based vulnerability detection methods~\cite{chakraborty2021deep, li2021vuldeelocator, fu2022linevul} predominantly focus on single-language scenarios.
Their design paradigms and technical architectures often lack the flexibility required to adapt to multi-language programming environments.
While some recent studies~\cite{fu2023chatgpt, purba2023software} have begun to assess the effectiveness of LLMs in vulnerability detection, these efforts are typically limited to small datasets and single-language settings, leaving the multi-language context largely unexplored.
Furthermore, it is crucial to distinguish between PLMs and LLMs in terms of their underlying architectures and intended applications.
PLMs are typically encoder-based models employed in classification tasks, where they learn discriminative features from labeled data to identify vulnerabilities.
In contrast, LLMs are generative, decoder-only models that leverage autoregressive mechanisms to generate the next token based on prior input, enabling them to produce vulnerability detection outputs in a more flexible manner.
Given these architectural and functional differences—as well as the current limitations of existing approaches—there is significant theoretical and practical value in conducting empirical research to systematically evaluate the effectiveness of automated vulnerability detection methods in multi-language settings.

In this study, we perform a preliminary study to empirically investigate the capabilities of both PLMs and LLMs for multilingual vulnerability detection.
Our experiments utilize a comprehensive dataset containing 20,165 code samples across seven programming languages (C, C\#, C++, Go, JavaScript, Java, and Python).
Through rigorous comparative analysis, we evaluate and contrast the performance of five representative PLMs and seven state-of-the-art LLMs on the vulnerability detection task, examining their effectiveness across different programming languages and in identifying top-ranking vulnerabilities.

This study presents the first attempt to evaluate and compare the performance of PLMs and LLMs for multilingual vulnerability detection.
Through a comprehensive large-scale benchmarking analysis, we demonstrate that the PLM CodeT5P achieves superior performance across most evaluation metrics, while existing LLMs exhibit promising but inconsistent results, indicating a need for further improvement.
We further conduct an in-depth analysis of model effectiveness in detecting the top 25 most dangerous software weaknesses.
Our findings reveal that CodeT5P consistently outperforms other LLMs in identifying real-world critical vulnerabilities, whereas closed-source LLMs exhibit stronger detection capabilities compared to their open-source counterparts.
These findings provide actionable guidance for selecting and optimizing language models across diverse programming environments, thereby advancing research in automated vulnerability detection.

%% file: content/02-background.tex
\section{Related Work}
\textbf{Software Vulnerability} is a defect or weakness present in software systems that attackers can exploit to launch attacks on the software~\cite{nist_csrc_vulnerability}.  
The impact can be severe, as unpatched vulnerabilities in software ecosystems may lead to catastrophic consequences, including significant economic losses~\cite{bilge12empirical}.  
To classify different types of vulnerability instances, Common Weakness Enumeration (CWE) and Common Vulnerabilities and Exposures (CVE) are used to refer to software defect types that may lead to vulnerabilities and specific instances of vulnerabilities in software systems, respectively.  
Software vulnerabilities are frequently discovered in open-source projects.  
~\citet{alfadel2023empirical} provided in-depth insights into common security issues in the Python ecosystem by analyzing 550 vulnerability reports affecting 252 Python packages.  
~\citet{hu2024empirical} investigated the prevalence and remediation delays of vulnerabilities in Go modules, finding that 66.10\% of modules were affected and identifying two types of delays hindering timely vulnerability fixes.  
~\citet{meng2018secure} examined the challenges developers face in implementing secure Java coding practices, highlighting vulnerabilities such as insecure configurations and misuse of security libraries, while proposing recommendations to improve the security of Java applications.
\citet{zerouali2022impact} investigated the impact of vulnerable software packages on downstream code repositories and found that nearly two-thirds of JavaScript and Ruby projects indirectly rely on vulnerable npm and RubyGems packages.
Through analyzing the PyPI and Maven ecosystems, \citet{xu2022insight} identified over 80,000 Python and Java projects with direct or indirect dependencies on vulnerable C software packages.

\smallskip
\noindent
\textbf{Automated Vulnerability Detection} is a security assessment method that automatically identifies potential security weaknesses in software or systems through techniques such as static code analysis~\cite{10.1145/3576915.3624401} or dynamic execution testing~\cite{kim2019rvfuzzer}.  
In recent years, groundbreaking advancements in deep learning have significantly driven innovative applications of automated vulnerability detection methods~\cite{shiri2024systematic}.  
These methods learn potential vulnerability patterns by constructing abstract representations of source code and establishing a nonlinear mapping relationship between source code representations and vulnerability existence (i.e., whether a given code snippet contains security vulnerabilities).  
ReVeal~\cite{chakraborty2021deep} employs the SMOTE method~\cite{chawla2002smote} to address data imbalance and uses a triplet loss function to maximize the separation between vulnerable and non-vulnerable code.   
VulDeeLocator~\cite{li2021vuldeelocator} captures semantic information in source code by introducing intermediate code and adopts a progressive refinement strategy, narrowing down the vulnerability localization scope progressively from coarse to fine granularity, thereby achieving high-precision vulnerability detection and localization.   
LineVul~\cite{fu2022linevul} employs a CodeBERT~\cite{feng2020codebert} model to generate code vector representations, effectively capturing lexical and semantic features for enhanced embeddings.
By leveraging BERT’s attention mechanism~\cite{devlin2019bert}, it precisely localizes vulnerable lines, enabling finer-grained detection than traditional methods.
\citet{fu2023chatgpt} examined ChatGPT's effectiveness in identifying C/C++ vulnerabilities.
These studies~\cite{chakraborty2021deep, li2021vuldeelocator, fu2022linevul, fu2023chatgpt} have established a crucial foundation for deep learning-based vulnerability detection, significantly advancing the field.
While existing methods demonstrate remarkable performance in monolingual settings, their adaptability to multilingual vulnerability contexts, particularly for PLMs and LLMs, has yet to be explored.

%% file: content/03-evaluation.tex
\section{Study Design}

Building upon the recently released multilingual vulnerability dataset REEF ~\cite{wang2023reef}, we construct a comprehensive benchmark for multilingual vulnerability detection.
Our study systematically investigates three key aspects of automated vulnerability detection in multilingual settings. 
First, we evaluate the effectiveness of widely used PLMs and LLMs for identifying vulnerabilities across languages.
Next, we examine how detection performance varies across different programming languages to uncover potential disparities.
Finally, through orthogonal analysis, we characterize the strengths and limitations of existing AVD techniques, providing insights into their complementary capabilities.

\subsection{Dataset Preparation}

\begin{table}[htbp]
\tiny
\caption{Statistical summary of multilingual vulnerability detection dataset}
\label{tab:dataset}
\centering
\begin{adjustbox}{width=0.45\textwidth,center}
\begin{threeparttable}
\begin{tabular}{ccccc}
\toprule
\textbf{Language} & \textbf{Training Set} & \textbf{Validation Set} & \textbf{Test Set} & \textbf{Total} \\
\midrule
C & 2,444 & 305 & 307 & 3,056 \\
C++ & 1,432 & 179 & 181 & 1,792 \\
C\# & 341 & 42 & 44 & 427 \\
Go & 2,323 & 290 & 292 & 2,905 \\
Java & 2,587 & 323 & 325 & 3,235 \\
JavaScript & 4,374 & 546 & 548 & 5,468 \\
Python & 2,625 & 328 & 329 & 3,282 \\
\midrule
Total & 16,126 & 2,013 & 2,026 & 20,165 \\
\bottomrule
\end{tabular}
\end{threeparttable}
\end{adjustbox}
\end{table}

\textbf{Studied Subject.}
To evaluate the performance of automated vulnerability detection techniques, we select the multilingual vulnerability dataset REEF ~\cite{wang2023reef}, which contains 4,466 CVEs and 30,987 patches (covering 236 CWEs) across seven programming languages.  
Specifically, REEF collects real-world vulnerabilities from the National Vulnerability Database~\cite{nvd} (NVD) and the CVE list maintained by Mend ~\cite{whitesource2022mend}, an open-source vulnerability database that aggregates CVEs from 2016 to 2023.  
In all, the paper focuses on all seven popular programming languages: C, C++, C\#, Go, Java, JavaScript, and Python.

\textbf{Data Preprocessing}.
To adapt the REEF dataset for automated vulnerability detection, we processed the data as follows:
First, we generated pre-change and post-change code versions using the Linux patch command\footnote{\url{https://www.man7.org/linux/man-pages/man1/patch.1.html}}.
Next, we removed comments and extracted function definitions using Tree-sitter~\cite{brunsfeld2024tree}.
We matched functions between the two versions based on their names, and for cases with multiple matches, we selected the closest pair using edit distance.
Finally, each pre-change function was labeled as a vulnerable (positive) sample, while each post-change function was labeled as a non-vulnerable (negative) sample.
Considering the input limitations of existing vulnerability detection methods, only data with no more than 512 tokens in length are retained.
Following existing work~\cite{fu2022linevul}, we adopted a widely used random sampling strategy, the dataset is split into training, validation, and test sets in an 8:1:1 ratio.  
To ensure that data for different languages maintains a consistent distribution ratio, we first allocated data for each language according to predetermined proportions.  
Subsequently, the corresponding portions are assembled to form the final dataset.  
As shown in Table~\ref{tab:dataset}, the vulnerability detection dataset contains 16,126 training samples, 2,013 validation samples, and 2,026 test samples, covering seven programming languages.

\subsection{Pre-trained Language Models}
For pre-trained language models, we select a variety of models, which are widely used in the literature~\cite{fu2022linevul, tian2024large}, including:  
\begin{itemize}  
    \item \textbf{UniXcoder}~\cite{guo2022unixcoder} is a unified cross-modal pre-trained language model.
    It enhances code representation by integrating multimodal data such as abstract syntax trees and code comments and employs a masked attention matrix mechanism along with a prefix adapter structure to achieve precise control over model behavior.    
    \item \textbf{CodeBERT}~\cite{feng2020codebert} is a widely recognized pre-trained language model that leverages a multi-layer Transformer architecture to effectively learn from bimodal data, including source code and natural language.  
    \item \textbf{CodeT5}~\cite{wang2021codet5} is a unified encoder-decoder model that improves performance by incorporating token type information from code, extending the T5 architecture.  
    \item \textbf{CodeT5P}~\cite{wang2023codet5+} is an improvement over CodeT5.
    It adopts a shallow encoder and deep decoder and is trained in multiple stages, initially using unimodal data and later incorporating bimodal data.  
    \item \textbf{LineVul}~\cite{fu2022linevul} achieves precise localization of vulnerable code lines through the attention mechanism in the BERT architecture, providing finer-grained vulnerability detection capabilities compared to traditional methods.  
\end{itemize}

\subsection{Large Language Models}  
\label{subsec:study_llm}  

This paper investigates seven advanced large language models (including three open-source LLMs and four closed-source LLMs), which have demonstrated outstanding performance in various code-related downstream tasks~\cite{wang2024software}.
Detailed information about the LLMs used in this paper is provided below:  

\begin{itemize}  
    \item \textbf{DeepSeek-Coder}~\cite{guo2024deepseek} was trained on 2 trillion tokens across 87 programming languages, leveraging a 16K context window and fill-in-the-blank tasks to enhance code generation and infilling capabilities. 
    It achieves state-of-the-art performance among open-source models and surpasses closed-source models such as GPT-3.5 in certain tasks.  
    \item \textbf{Code Llama}~\cite{roziere2023code} is a decoder-only model and one of the most popular LLMs for code generation and infilling, built upon the Llama 2~\cite{touvron2023llama} model.
    It was further fine-tuned on 500 billion tokens extracted from a vast and diverse code dataset.  

    \item \textbf{Llama 3.1}~\cite{dubey2024llama} is a series of large language models based on the Transformer architecture, designed to improve performance across various language understanding tasks.
    By optimizing data quality, training scale, and model architecture, Llama 3.1 has demonstrated significant potential in natural language processing.  
      
    \item \textbf{ChatGPT}~\cite{chatgpt2022} is a groundbreaking LLM capable of transforming various domains with its advanced natural language processing capabilities.
    It was trained on a massive corpus of natural language text and code snippets, employing reinforcement learning (RL) to enhance its ability to follow human instructions.
    Specifically, we examine two LLMs: GPT-3.5-Turbo~\cite{chatgpt2022} and GPT-4o~\cite{openai2024gpt4o}.

    \item\textbf{DeepSeek-R1}~\cite{guo2025deepseek} extensively employed reinforcement learning techniques during the post-training phase, significantly enhancing the model's reasoning capabilities with only minimal annotated data.
    In tasks such as mathematics, coding, and natural language reasoning, its performance rivals that of OpenAI o1~\cite{openai_o1_2024}.

    \item \textbf{QwQ-plus}~\cite{aliyun_qwq_2024} extends the Qwen2.5 architecture~\cite{yang2024qwen2} with reinforcement learning to significantly enhance inference capabilities.
\end{itemize}

\begin{figure}[ht]
\centering
\includegraphics[width=1.\linewidth]{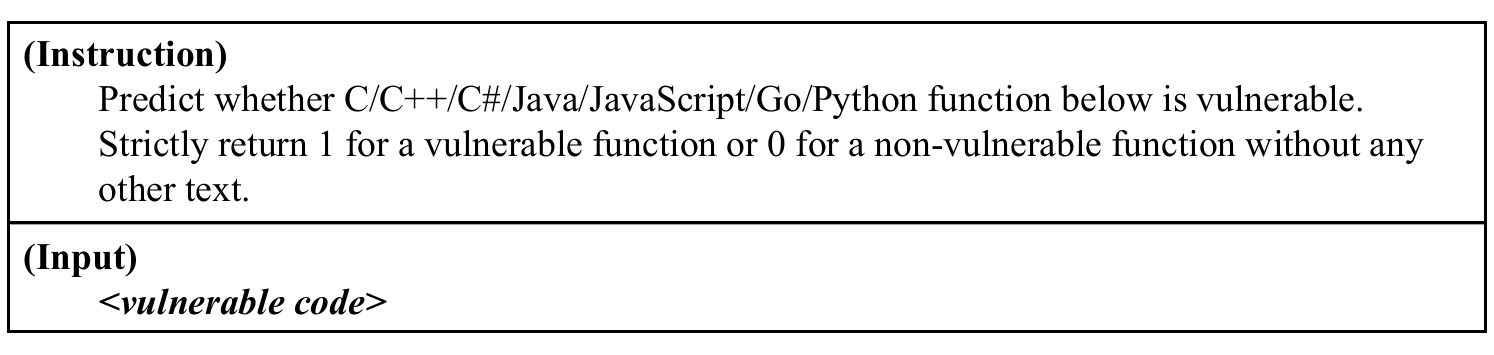}
\caption{Template of zero-shot prompting strategy for automated vulnerability detection}
\label{fig:study_avd_func_zsp}
\end{figure}

As a preliminary study, we design a zero-shot prompting strategy for automated vulnerability detection.
As shown in Figure~\ref{fig:study_avd_func_zsp}, we directly instruct the large language models to determine whether a given code contains vulnerabilities.

\subsection{Evaluation Metrics}

Automated vulnerability detection is a fundamental binary classification task.
To thoroughly assess the performance of vulnerability detection approaches, we adopt four widely-used evaluation metrics following previous work~\cite{fu2022linevul}: accuracy, recall, precision, and F1-Score.
Accuracy is the ratio of correct predictions to total predictions, ideal for balanced datasets.
Its calculation formula is:  
\begin{equation}  
\text{Accuracy} = \frac{TP + TN}{TP + TN + FP + FN}  
\end{equation}  
where \(TP\) (True Positive) represents true positives, \(TN\) (True Negative) represents true negatives, \(FP\) (False Positive) represents false positives, and \(FN\) (False Negative) represents false negatives.  
Recall checks if the model finds all true positives.
Its calculation formula is:  
\begin{equation}  
\text{Recall} = \frac{TP}{TP + FN}  
\end{equation}  
Precision reflects the model's correctness when it predicts a positive case.
Its calculation formula is:  
\begin{equation}  
\text{Precision} = \frac{TP}{TP + FP}  
\end{equation}  
F1-Score harmonizes precision and recall into one measure, ideal for uneven class distributions.
Its calculation formula is:  
\begin{equation}  
\text{F1-Score} = 2 \cdot \frac{\text{Precision} \cdot \text{Recall}}{\text{Precision} + \text{Recall}}  
\end{equation}

\subsection{Implementation and Environment}  
\label{subsec:implementation}  
We reproduced LineVul based on its implementation documentation and the parameter settings provided in the reproduction package.  
All open-source pre-trained language models (i.e., UniXcoder, CodeBERT, CodeT5, and CodeT5P) were downloaded from Huggingface~\cite{wolf2019huggingface}, adopting the parameter settings recommended in this work~\cite{fu2022linevul}.
For closed-source LLMs, we utilized OpenAI's API~\cite{openai2024} to access GPT-3.5-Turbo and GPT-4o.
Specifically, we employed gpt-3.5-turbo-0125 as the implementation version for GPT-3.5-Turbo and GPT-4-0613 as the representative version for GPT-4o in our experiments.
Additionally, we incorporated DeepSeek-R1 via DeepSeek's API~\cite{deepseek2024api} and QwQ-plus through Alibaba Cloud's API~\cite{aliyun_qwq_2024} to ensure a comprehensive comparison across different model architectures and providers.
For open-source LLMs (i.e., DeepSeek-Coder, Code Llama, and Llama 3.1), the pre-trained language models were downloaded from Huggingface~\cite{wolf2019huggingface}.  
In particular, the model sizes of these three open-source LLMs were 6.7B, 7B, and 8B, respectively.  
All experiments were conducted on a machine equipped with an Intel Xeon CPU Gold-6342, 512 GB of RAM, Ubuntu 20.04.6, and two A800 GPUs.

%% file: content/04-result.tex
\section{Results}

\subsection{RQ1: How effective are PLMs and LLMs for multilingual vulnerability detection?}
\label{sec:study_rq1}
\noindent
\textbf{\underline{Approach.}}
This research question aims to comparatively analyze the performance of various PLMs and LLMs in detecting multilingual security vulnerabilities.
Specifically, we investigate the effectiveness of five PLMs and seven LLMs for vulnerability detection.
Below we elaborate on the training process.
For vulnerability detection, given a function as input, the model predicts whether it is vulnerable.
For encoder-only models (CodeBERT and UniXcoder), we append a multilayer perceptron~\cite{kruse2022multi} (MLP)-based classifier and fine-tune the entire architecture for vulnerability detection.
Other models are adapted through direct fine-tuning on the REEF dataset.
For LLM-based approaches, we employ a direct generation method using a zero-shot prompting strategy, where LLMs perform end-to-end analysis of input code to directly assess potential vulnerabilities.

\begin{table}[htbp]
    \caption{Performance of PLMs and LLMs on multilingual vulnerability detection}
    \label{tab:study_rq1_func}
    \centering
    \small
    \begin{adjustbox}{max width=.95\textwidth, center}
    \begin{tabular}{lrrrr}
        \toprule
        \toprule
        \textbf{Technique} & \textbf{Accuracy} & \textbf{Recall} & \textbf{Precision} & \textbf{F1-Score}  \\ 
        \midrule
        \rowcolor{gray}\multicolumn{5}{l}{\textbf{PLM}} \\ 
        \midrule
            CodeBERT & 0.5163 & \textbf{1.0000} & 0.5098 & 0.6753 \\
            LineVul & 0.5173 & \textbf{1.0000} & 0.5103 & 0.6757 \\
            UniXcoder & 0.5814 & 0.8930 & 0.5518 & 0.6822 \\
            CodeT5 & 0.5854 & 0.9342 & 0.5519 & 0.6939 \\
            CodeT5P & \textbf{0.6037} & 0.9529 & \textbf{0.5626} & \textbf{0.7075} \\
        \midrule
        \rowcolor{gray}\multicolumn{5}{l}{\textbf{LLM}} \\ 
        \midrule
            DeepSeek-Coder &0.5005 &0.1796 &0.5097 & 0.2656  \\
            Code Llama & 0.4877 & 0.9156 & 0.4950 & 0.6426 \\
            Llama 3.1 & 0.5005 & 0.5015 & 0.5034 & 0.5025 \\
            GPT-3.5-Turbo & 0.4758 & 0.6055 & 0.4832 & 0.5375 \\
            GPT-4o & 0.4970  & 0.2601  & 0.5000  & 0.3422  \\
            DeepSeek-R1	 & 0.5217  & 	0.6585  & 	0.5193  & 	0.5807 \\
            QwQ-plus & 	0.5346  & 	0.4838  & 	0.5418  & 	0.5111 \\
        \bottomrule
        \bottomrule
    \end{tabular}
    \end{adjustbox}
\end{table}

\noindent
\textbf{\underline{Results.}}  
Table~\ref{tab:study_rq1_func} presents a comparison of accuracy, recall, precision, and F1-Score between PLMs and LLMs in multilingual vulnerability detection.
The experimental results demonstrate substantial performance variations across different PLMs and LLMs.
The bolded values indicate the methods that achieve optimal performance for each respective metric.

Among all evaluated models, the encoder-decoder-based CodeT5P achieves the best overall performance, attaining the highest scores in accuracy (0.6037), precision (0.5626), and F1-score (0.7075).
Notably, while CodeBERT and LineVul achieve perfect recall (1.0000), their precision remains relatively low (approximately 0.5000), indicating these models tend to produce a high rate of false positives.
Although encoder-only models demonstrate strong recall performance, the encoder-decoder architecture employed by CodeT5 and CodeT5P consistently outperforms all other PLMs and LLMs across the key metrics of accuracy, precision, and F1-score. 

Compared to PLMs, LLMs generally exhibit inferior performance in vulnerability detection. 
Among the evaluated LLMs, QwQ-plus achieves the highest accuracy (0.5346) but demonstrates relatively lower recall (0.4838), suggesting a tendency toward conservative vulnerability identification.
Notably, both QwQ-plus and DeepSeek-R1 outperform GPT-3.5-Turbo and GPT-4o, implying that their reinforcement learning-enhanced reasoning capabilities may improve task effectiveness.
For open-source LLMs, Llama 3.1 and DeepSeek-Coder attain identical accuracy scores (0.5005), though Llama 3.1 surpasses DeepSeek-Coder in other metrics.
Overall, current LLMs lag behind PLMs across nearly all evaluation metrics, highlighting the need for further research to optimize their applicability for vulnerability detection with effective prompting techniques.

\begin{tcolorbox}[boxsep=3pt]
\textbf{RQ1 Summary:}
CodeT5P demonstrates superior performance in vulnerability detection, achieving the highest scores in accuracy, precision, and F1-score among all evaluated models.
While current LLMs generally underperform compared to specialized PLMs, their emerging capabilities suggest significant potential that merits further investigation for vulnerability detection.
\end{tcolorbox}

\subsection{RQ2: How do PLMs and LLMs perform across different programming language vulnerabilities?}
\begin{figure}[htbp]
    \centering
    \includegraphics[width=1.\linewidth]{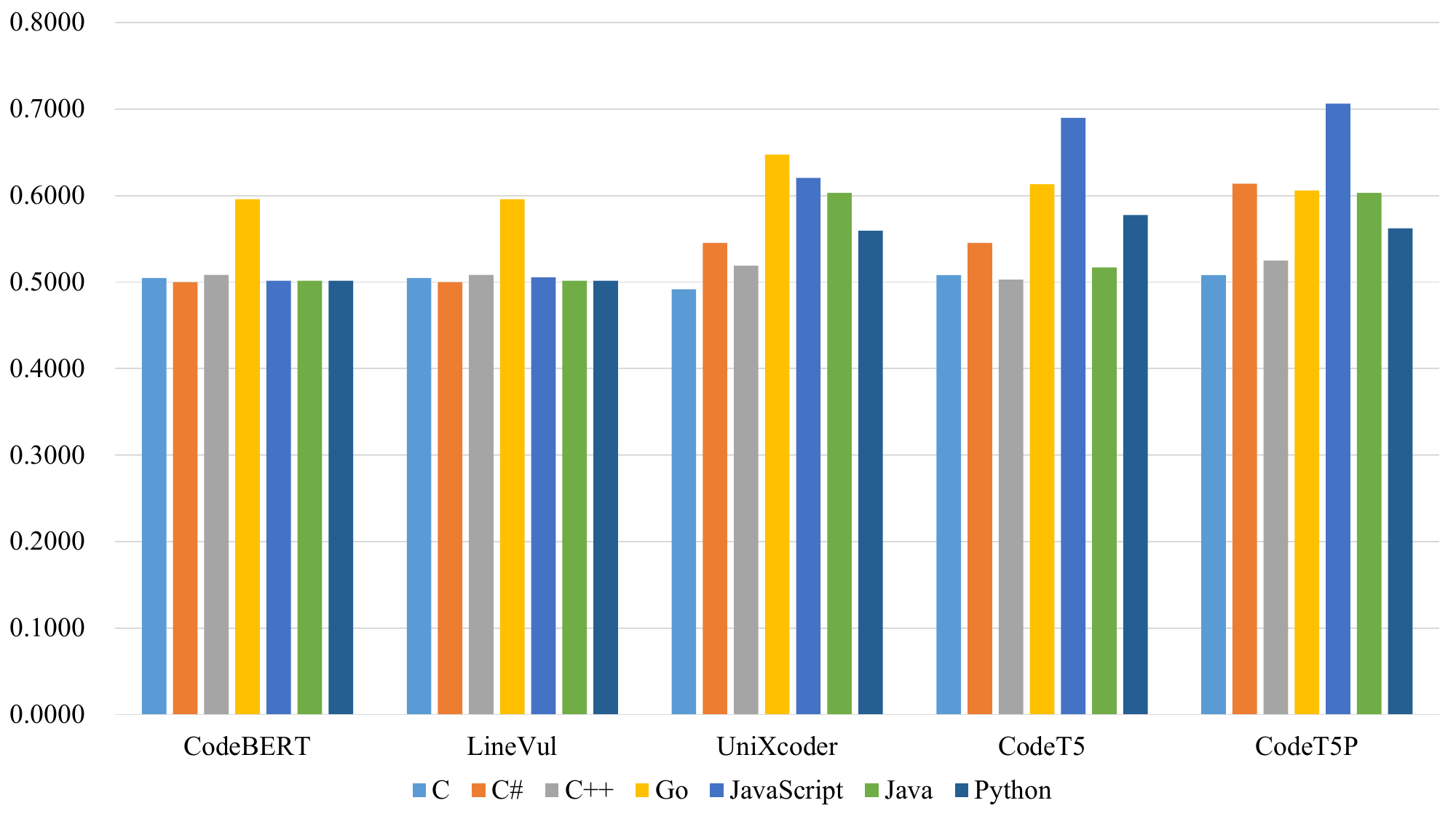}
    \caption{Performance of PLM techniques on vulnerability detection across seven languages (y-axis: accuracy)}
    \label{fig:study_rq2_func_plm}
\end{figure}

\noindent
\textbf{\underline{Approach.}}
Based on the empirical findings from RQ1 (Section~\ref{sec:study_rq1}), this research question conducts a comprehensive evaluation and comparative analysis of PLMs' and LLMs' performance across diverse programming languages.
We systematically examine their applicability, effectiveness, and performance variations, with a particular focus on detection accuracy across seven programming languages (selected due to the balanced distribution of positive and negative examples in the dataset).
This in-depth investigation not only provides theoretical foundations and practical insights for methodological optimization, but also establishes a critical basis for advancing multilingual vulnerability detection research.

\begin{figure}[htbp]
\centering
\includegraphics[width=1.\linewidth]{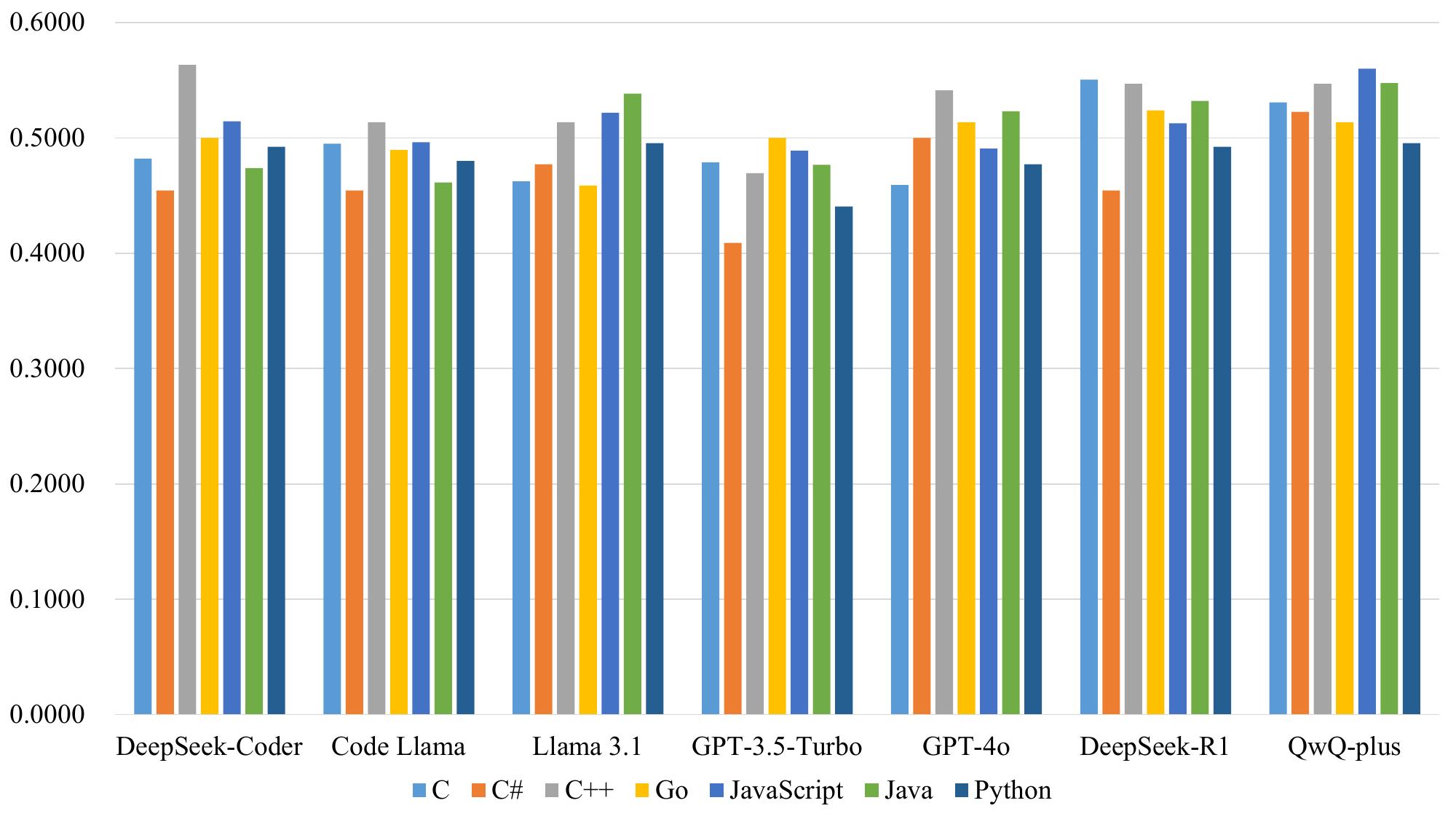}
\caption{Performance of LLM techniques on vulnerability detection across seven languages (y-axis: accuracy)}
\label{fig:study_rq2_func_llm}
\end{figure}

\noindent
\textbf{\underline{Results.}}  
Figures~\ref{fig:study_rq2_func_plm} and~\ref{fig:study_rq2_func_llm} present the performance evaluations of five PLMs and seven LLMs, respectively, for vulnerability detection across seven programming languages.

The experimental analysis yields several critical observations.
First, CodeBERT and LineVul exhibit exceptional proficiency in Go (both 0.5959 accuracy), while UniXcoder establishes state-of-the-art performance among PLMs for Go (0.6473 accuracy), indicating effective capture of language-specific vulnerability patterns.
Notably,  all PLMs do well in Go language among seven programming languages.
This may be attributed to the relatively recent emergence of the Go programming language.

Second, encoder-decoder models demonstrate optimal performance across all languages except Go among PLMs. CodeT5 achieves peak accuracy in C (0.5081) and Python (0.5775), while CodeT5P excels in C (0.5081), C\# (0.6136), C++ (0.5249), JavaScript (0.7062), and Java (0.6031).
These results substantiate the architectural advantages of encoder-decoder frameworks for multilingual vulnerability detection, suggesting promising research directions.

As revealed in Figure~\ref{fig:study_rq2_func_llm}, significant cross-linguistic performance variations emerge among LLMs.
DeepSeek-Coder achieves the highest accuracy in C++ (0.5635), while DeepSeek-R1 exhibits superior detection capabilities for C (0.5505) and Go (0.5240).
In comparison, QwQ-plus shows leading performance in C\# (0.5227), JavaScript (0.5602), Java (0.5477), and Python (0.4954).
Particularly, Llama 3.1 performs comparably to QwQ-plus in Python, achieving similar accuracy.
Notably, among both closed-source and open-source LLMs, the best-performing closed-source LLM consistently outperforms the top open-source counterpart across six programming languages, with the sole exception of DeepSeek-Coder's performance in C++. This observation suggests that capability improvements are significantly influenced by model scale.
The observed language-specific specialization among closed-source models emphasizes the necessity of aligning model selection with programming language characteristics for optimal zero-shot vulnerability detection.

Interestingly, DeepSeek-R1's superior performance in C (0.5505) and DeepSeek-Coder's advantage in C++ (0.5635) among all PLMs and LLMs collectively demonstrate the potential of LLMs in vulnerability detection.
These findings underscore the importance of strategic model-language pairing and consideration of architectural paradigms for effective vulnerability identification.

\begin{tcolorbox}[boxsep=3pt,]
\textbf{RQ2 Summary:}
Among PLMs, encoder-decoder architectures (particularly CodeT5P) achieve state-of-the-art performance across most languages, while LLMs (notably DeepSeek variants) exhibit superior capability in C/C++ vulnerability detection.
\end{tcolorbox}

\subsection{RQ3: How do PLMs and LLMs perform in detecting serious vulnerabilities?}
\label{subsec:RQ3}
\noindent
\textbf{\underline{Approach.}} 
This research question investigates the performance characteristics of various PLM and LLM techniques in detecting serious vulnerabilities.
Based on the findings from RQ1, we selected the top-performing techniques from each category:

\begin{itemize}
    \item \textbf{PLM technique:} CodeT5P
    \item \textbf{Open-source LLM technique:} Llama 3.1
    \item \textbf{Closed-source LLM techniques:} GPT-4o and QwQ-plus
\end{itemize}

To ensure a balanced comparison, we included both RL-enhanced and non-RL variants of closed-source LLMs.
Specifically, while QwQ-plus incorporates RL-based optimization, GPT-4o does not rely on RL fine-tuning.
We assess detection capabilities against the 2024 CWE Top 25 Most Dangerous Software Weaknesses\footnote{\url{https://cwe.mitre.org/top25/archive/2024/2024_cwe_top25.html}}—a prioritized list of severe vulnerabilities published by the MITRE Corporation.
Note that our test dataset covers all 25 CWE-IDs, enabling a rigorous evaluation of each technique’s effectiveness in identifying high-risk vulnerabilities.

\noindent
\textbf{\underline{Results.}} 
Table~\ref{tab:rq3_top25} shows the detection performance of studied AVD techniques on the top 25 most dangerous CWE-IDs in 2024 in terms of accuracy and the values highlighted in bold denote the technique that exhibits the best performance for corresponding CWE-ID.

\begin{table*}[t]
    \caption{The accuracy of four techniques across the Top 25 Most Dangerous CWE-IDs in 2024}
    \label{tab:rq3_top25}
    \centering
    \small
    \begin{adjustbox}{max width=1.0\textwidth, center}
    \begin{threeparttable}
    \begin{tabular}{ccrrrrc}
        \toprule
        \toprule
        Rank & CWE-ID & CodeT5P & Llama 3.1 & GPT-4o& QwQ-plus  & Total \\
        \midrule
        1  & CWE-79                                                  & \textbf{125(63.13\%)} & 101(51.01\%)         & 101(51.01\%)         & 118(59.6\%)          & 198          \\
        2  & CWE-787                                                   & \textbf{23(56.1\%)}   & 18(43.9\%)           & 21(51.22\%)          & \textbf{23(56.1\%)}  & 41           \\
        3  & CWE-89                                                  & 27(58.7\%)            & 15(32.61\%)          & 25(54.35\%)          & \textbf{30(65.22\%)} & 46           \\
        4  & CWE-352                                              & \textbf{62(77.5\%)}   & 44(55.0\%)           & 37(46.25\%)          & 46(57.5\%)           & 80           \\
        5  & CWE-22                                                          & 43(56.58\%)           & 36(47.37\%)          & 43(56.58\%)          & \textbf{48(63.16\%)} & 76           \\
        6  & CWE-125                                                       & \textbf{31(60.78\%)}  & 23(45.1\%)           & 21(41.18\%)          & 30(58.82\%)          & 51           \\
        7  & CWE-78                                                     & 5(33.33\%)            & 3(20.0\%)            & \textbf{7(46.67\%)}  & \textbf{7(46.67\%)}  & 15           \\
        8  & CWE-416                                                    & \textbf{19(55.88\%)}  & 14(41.18\%)          & 16(47.06\%)          & \textbf{19(55.88\%)} & 34           \\
        9  & CWE-862                                              & \textbf{1(100.0\%)}   & 0(0.0\%)             & 0(0.0\%)             & 0(0.0\%)             & 1            \\
        10 & CWE-434                         & 3(60.0\%)             & 2(40.0\%)            & \textbf{4(80.0\%)}   & 2(40.0\%)            & 5            \\
        11 & CWE-94                                                   & \textbf{18(75.0\%)}   & 10(41.67\%)          & 12(50.0\%)           & 14(58.33\%)          & 24           \\
        12 & CWE-20                                         & \textbf{52(56.52\%)}  & 37(40.22\%)          & 39(42.39\%)          & 41(44.57\%)          & 92           \\
        13 & CWE-77                                               & 14(53.85\%)           & 11(42.31\%)          & \textbf{16(61.54\%)} & 14(53.85\%)          & 26           \\
        14 & CWE-287                                             & \textbf{110(78.57\%)} & 65(46.43\%)          & 67(47.86\%)          & 71(50.71\%)          & 140          \\
        15 & CWE-269                                   & \textbf{8(72.73\%)}   & 5(45.45\%)           & 5(45.45\%)           & 5(45.45\%)           & 11           \\
        16 & CWE-502                        & \textbf{18(62.07\%)}  & 16(55.17\%)          & 17(58.62\%)          & 15(51.72\%)          & 29           \\
        17 & CWE-200       & 37(49.33\%)           & \textbf{43(57.33\%)} & 42(56.0\%)           & 40(53.33\%)          & 75           \\
        18 & CWE-863                             & 22(51.16\%)           & 22(51.16\%)          & \textbf{23(53.49\%)} & 19(44.19\%)          & 43           \\
        19 & CWE-918                                 & 20(60.61\%)           & \textbf{21(63.64\%)} & 15(45.45\%)          & 17(51.52\%)          & 33           \\
        20 & CWE-119   & 30(48.39\%)           & 31(50.0\%)           & 32(51.61\%)          & \textbf{36(58.06\%)} & 62           \\
        21 & CWE-476                      & 19(42.22\%)           & 21(46.67\%)          & 22(48.89\%)          & \textbf{23(51.11\%)} & 45           \\
        22 & CWE-798                                  & 1(33.33\%)            & 0(0.0\%)             & \textbf{2(66.67\%)}  & 1(33.33\%)           & 3            \\
        23 & CWE-190                               & \textbf{18(56.25\%)}  & \textbf{18(56.25\%)} & 17(53.12\%)          & \textbf{18(56.25\%)} & 32           \\
        24 & CWE-400                           & \textbf{46(60.53\%)}  & 35(46.05\%)          & 38(50.0\%)           & 42(55.26\%)          & 76           \\
        25 & CWE-306         & \textbf{9(50.0\%)}    & 8(44.44\%)           & 7(38.89\%)           & \textbf{9(50.0\%)}   & 18          \\
                \midrule
         & Average &\textbf{ 761(60.59\%)} & 599(47.69\%) & 629(50.08\%) & 688(54.78\%) & 1256 \\
        \bottomrule
        \bottomrule
    \end{tabular}
    \begin{tablenotes}
    \small
        \item[*] The number in the parentheses represents the corresponding accuracy of techniques, and Total column represents the total amount of data for corresponding CWE-ID in our test dataset.
    \end{tablenotes}
    
    \end{threeparttable}
    \end{adjustbox}
    
\end{table*}

Our analysis of Table~\ref{tab:rq3_top25} reveals that CodeT5P demonstrates superior vulnerability detection capabilities among the evaluated techniques.
Specifically, it successfully identifies 761 out of 1256 functions (60.59\%) associated with the top 25 most dangerous CWE-IDs in 2024, outperforming other models by significant margins: 27.05\% higher than Llama 3.1 (599 detections), 20.99\% higher than GPT-4o (629 detections), and 10.61\% higher than QwQ-plus (688 detections). 
A detailed breakdown shows CodeT5P achieves the best performance on 14 out of the 25 CWE-IDs, while Llama 3.1, GPT-4o, and QwQ-plus lead on only 3, 5, and 9 CWE-IDs respectively.
Notably, CodeT5P uniquely demonstrates the capability to detect CWE-862 vulnerabilities, which other models fail to identify.

These results highlight two key findings: First, PLMs, particularly CodeT5P, exhibit remarkable efficacy in detecting critical software vulnerabilities.
Second, the observed performance gap between closed-source LLMs (e.g., GPT-4o, QwQ-plus) and open-source alternatives (e.g., Llama 3.1) may be partially attributable to model scale.
For instance, while GPT-4o and QwQ-plus successfully detect CWE-798 vulnerabilities, Llama 3.1 fails to identify them.
This observation suggests that future research should investigate whether scaling open-source models beyond Llama 3.1's 8B parameters could bridge this performance gap in the vulnerability detection task.

\begin{tcolorbox}\textbf{RQ3 Summary:}
CodeT5P excels in vulnerability detection, demonstrating superior performance on high-risk CWE-IDs.
Meanwhile, closed-source LLMs consistently outperform open-source models across these critical vulnerabilities.
\end{tcolorbox}

%% file: content/05-discussion.tex
\section{Future Work}
Future research directions could explore vulnerability detection at multiple granularities, including both file-level and line-level analysis, within the context of multilingual vulnerability detection task.
Furthermore, while the current study investigates LLMs using only a zero-shot prompting approach, subsequent work could examine alternative prompt strategies such as few-shot prompting, instruction-tuning, and RAG techniques.
Additionally, due to computational constraints, the present study is limited to open-source LLMs with approximately 8 billion parameters;
future investigations could extend this work to larger-scale language models with significantly more parameters.

%% file: content/06-threat.tex
\section{Threats to Validity}
This study has certain limitations that should be acknowledged. First, since our work relies on the REEF dataset, which covers seven widely used programming languages, the generalizability of our findings to other languages may be limited.
For future work, we plan to expand our collection of vulnerabilities across a broader range of programming languages using the REEF framework.
Following previous work, we exclude functions longer than 512 tokens due to the intrinsic limitations of PLMs, which may restrict the generalizability of our findings to longer functions.
To address this potential limitation, we will explore techniques that could extend the input length of PLMs, such as Fusion-in-Decoder~\cite{izacard-grave-2021-leveraging}.
To address potential internal validity threats stemming from our implementation of the automated detection approaches, we adopted rigorous mitigation strategies. 
Specifically, we faithfully reimplemented all methods using the replication packages provided in their original papers. 
Furthermore, two authors independently reviewed the source code to ensure correctness and consistency with the established methodologies.
For performance evaluation, we employed well-established metrics in vulnerability research, including accuracy, recall, precision, and F1-score, to ensure a comprehensive and reliable assessment of the approaches under study.

%% file: content/00-main.bbl

\begin{thebibliography}{45}


\ifx \showCODEN    \undefined \def \showCODEN     #1{\unskip}     \fi
\ifx \showISBNx    \undefined \def \showISBNx     #1{\unskip}     \fi
\ifx \showISBNxiii \undefined \def \showISBNxiii  #1{\unskip}     \fi
\ifx \showISSN     \undefined \def \showISSN      #1{\unskip}     \fi
\ifx \showLCCN     \undefined \def \showLCCN      #1{\unskip}     \fi
\ifx \shownote     \undefined \def \shownote      #1{#1}          \fi
\ifx \showarticletitle \undefined \def \showarticletitle #1{#1}   \fi
\ifx \showURL      \undefined \def \showURL       {\relax}        \fi
\providecommand\bibfield[2]{#2}
\providecommand\bibinfo[2]{#2}
\providecommand\natexlab[1]{#1}
\providecommand\showeprint[2][]{arXiv:#2}

\bibitem[Alfadel et~al\mbox{.}(2023)]%
        {alfadel2023empirical}
\bibfield{author}{\bibinfo{person}{Mahmoud Alfadel}, \bibinfo{person}{Diego~Elias Costa}, {and} \bibinfo{person}{Emad Shihab}.} \bibinfo{year}{2023}\natexlab{}.
\newblock \showarticletitle{Empirical analysis of security vulnerabilities in python packages}.
\newblock \bibinfo{journal}{\emph{Empirical Software Engineering}} \bibinfo{volume}{28}, \bibinfo{number}{3} (\bibinfo{year}{2023}), \bibinfo{pages}{59}.
\newblock


\bibitem[Bilge and Dumitras({[n.\,d.]})]%
        {bilge12empirical}
\bibfield{author}{\bibinfo{person}{Leyla Bilge} {and} \bibinfo{person}{Tudor Dumitras}.} \bibinfo{year}{[n.\,d.]}\natexlab{}.
\newblock \showarticletitle{An empirical study of zeroday attacks in the real world}.
\newblock \bibinfo{journal}{\emph{CCS’12}} (\bibinfo{year}{[n.\,d.]}), \bibinfo{pages}{16--18}.
\newblock


\bibitem[Brunsfeld(2024)]%
        {brunsfeld2024tree}
\bibfield{author}{\bibinfo{person}{Max Brunsfeld}.} \bibinfo{year}{2024}\natexlab{}.
\newblock \bibinfo{title}{tree-sitter/tree-sitter: v0.23.0}.
\newblock
\href{https://doi.org/10.5281/zenodo.13375512}{doi:\nolinkurl{10.5281/zenodo.13375512}}


\bibitem[Chakraborty et~al\mbox{.}(2021)]%
        {chakraborty2021deep}
\bibfield{author}{\bibinfo{person}{Saikat Chakraborty}, \bibinfo{person}{Rahul Krishna}, \bibinfo{person}{Yangruibo Ding}, {and} \bibinfo{person}{Baishakhi Ray}.} \bibinfo{year}{2021}\natexlab{}.
\newblock \showarticletitle{Deep learning based vulnerability detection: Are we there yet?}
\newblock \bibinfo{journal}{\emph{IEEE Transactions on Software Engineering}} \bibinfo{volume}{48}, \bibinfo{number}{9} (\bibinfo{year}{2021}), \bibinfo{pages}{3280--3296}.
\newblock


\bibitem[Chawla et~al\mbox{.}(2002)]%
        {chawla2002smote}
\bibfield{author}{\bibinfo{person}{Nitesh~V Chawla}, \bibinfo{person}{Kevin~W Bowyer}, \bibinfo{person}{Lawrence~O Hall}, {and} \bibinfo{person}{W~Philip Kegelmeyer}.} \bibinfo{year}{2002}\natexlab{}.
\newblock \showarticletitle{SMOTE: synthetic minority over-sampling technique}.
\newblock \bibinfo{journal}{\emph{Journal of artificial intelligence research}}  \bibinfo{volume}{16} (\bibinfo{year}{2002}), \bibinfo{pages}{321--357}.
\newblock


\bibitem[Cloud(2024)]%
        {aliyun_qwq_2024}
\bibfield{author}{\bibinfo{person}{Alibaba Cloud}.} \bibinfo{year}{2024}\natexlab{}.
\newblock \bibinfo{title}{QwQ Model Documentation}.
\newblock
\urldef\tempurl%
\url{https://www.alibabacloud.com/help/en/model-studio/user-guide/qwq}
\showURL{%
\tempurl}
\newblock
\shownote{Accessed: March 29, 2025}.


\bibitem[Dam et~al\mbox{.}(2018)]%
        {dam2018automatic}
\bibfield{author}{\bibinfo{person}{Hoa~Khanh Dam}, \bibinfo{person}{Truyen Tran}, \bibinfo{person}{Trang Pham}, \bibinfo{person}{Shien~Wee Ng}, \bibinfo{person}{John Grundy}, {and} \bibinfo{person}{Aditya Ghose}.} \bibinfo{year}{2018}\natexlab{}.
\newblock \showarticletitle{Automatic feature learning for predicting vulnerable software components}.
\newblock \bibinfo{journal}{\emph{IEEE Transactions on Software Engineering}} \bibinfo{volume}{47}, \bibinfo{number}{1} (\bibinfo{year}{2018}), \bibinfo{pages}{67--85}.
\newblock


\bibitem[DeepSeek(2024)]%
        {deepseek2024api}
\bibfield{author}{\bibinfo{person}{DeepSeek}.} \bibinfo{year}{2024}\natexlab{}.
\newblock \bibinfo{booktitle}{\emph{DeepSeek API Documentation: Reasoning Model Guide}}.
\newblock DeepSeek.
\newblock
\urldef\tempurl%
\url{https://api-docs.deepseek.com/guides/reasoning_model}
\showURL{%
\tempurl}


\bibitem[Devlin et~al\mbox{.}(2019)]%
        {devlin2019bert}
\bibfield{author}{\bibinfo{person}{Jacob Devlin}, \bibinfo{person}{Ming-Wei Chang}, \bibinfo{person}{Kenton Lee}, {and} \bibinfo{person}{Kristina Toutanova}.} \bibinfo{year}{2019}\natexlab{}.
\newblock \showarticletitle{Bert: Pre-training of deep bidirectional transformers for language understanding}. In \bibinfo{booktitle}{\emph{Proceedings of the 2019 conference of the North American chapter of the association for computational linguistics: human language technologies, volume 1 (long and short papers)}}. \bibinfo{pages}{4171--4186}.
\newblock


\bibitem[Dubey et~al\mbox{.}(2024)]%
        {dubey2024llama}
\bibfield{author}{\bibinfo{person}{Abhimanyu Dubey}, \bibinfo{person}{Abhinav Jauhri}, \bibinfo{person}{Abhinav Pandey}, \bibinfo{person}{Abhishek Kadian}, \bibinfo{person}{Ahmad Al-Dahle}, \bibinfo{person}{Aiesha Letman}, \bibinfo{person}{Akhil Mathur}, \bibinfo{person}{Alan Schelten}, \bibinfo{person}{Amy Yang}, \bibinfo{person}{Angela Fan}, {et~al\mbox{.}}} \bibinfo{year}{2024}\natexlab{}.
\newblock \showarticletitle{The llama 3 herd of models}.
\newblock \bibinfo{journal}{\emph{arXiv preprint arXiv:2407.21783}} (\bibinfo{year}{2024}).
\newblock


\bibitem[Feng et~al\mbox{.}(2020)]%
        {feng2020codebert}
\bibfield{author}{\bibinfo{person}{Zhangyin Feng}, \bibinfo{person}{Daya Guo}, \bibinfo{person}{Duyu Tang}, \bibinfo{person}{Nan Duan}, \bibinfo{person}{Xiaocheng Feng}, \bibinfo{person}{Ming Gong}, \bibinfo{person}{Linjun Shou}, \bibinfo{person}{Bing Qin}, \bibinfo{person}{Ting Liu}, \bibinfo{person}{Daxin Jiang}, {et~al\mbox{.}}} \bibinfo{year}{2020}\natexlab{}.
\newblock \showarticletitle{Codebert: A pre-trained model for programming and natural languages}.
\newblock \bibinfo{journal}{\emph{arXiv preprint arXiv:2002.08155}} (\bibinfo{year}{2020}).
\newblock


\bibitem[Fu and Tantithamthavorn(2022)]%
        {fu2022linevul}
\bibfield{author}{\bibinfo{person}{Michael Fu} {and} \bibinfo{person}{Chakkrit Tantithamthavorn}.} \bibinfo{year}{2022}\natexlab{}.
\newblock \showarticletitle{Linevul: A transformer-based line-level vulnerability prediction}. In \bibinfo{booktitle}{\emph{Proceedings of the 19th International Conference on Mining Software Repositories}}. \bibinfo{pages}{608--620}.
\newblock


\bibitem[Fu et~al\mbox{.}(2023)]%
        {fu2023chatgpt}
\bibfield{author}{\bibinfo{person}{Michael Fu}, \bibinfo{person}{Chakkrit~Kla Tantithamthavorn}, \bibinfo{person}{Van Nguyen}, {and} \bibinfo{person}{Trung Le}.} \bibinfo{year}{2023}\natexlab{}.
\newblock \showarticletitle{Chatgpt for vulnerability detection, classification, and repair: How far are we?}. In \bibinfo{booktitle}{\emph{2023 30th Asia-Pacific Software Engineering Conference (APSEC)}}. IEEE, \bibinfo{pages}{632--636}.
\newblock


\bibitem[Gobbi and Kinder(2023)]%
        {10.1145/3576915.3624401}
\bibfield{author}{\bibinfo{person}{Mat\'{\i}as~F. Gobbi} {and} \bibinfo{person}{Johannes Kinder}.} \bibinfo{year}{2023}\natexlab{}.
\newblock \showarticletitle{Poster: Using CodeQL to Detect Malware in npm}. In \bibinfo{booktitle}{\emph{Proceedings of the 2023 ACM SIGSAC Conference on Computer and Communications Security}} (Copenhagen, Denmark) \emph{(\bibinfo{series}{CCS '23})}. \bibinfo{publisher}{Association for Computing Machinery}, \bibinfo{address}{New York, NY, USA}, \bibinfo{pages}{3519–3521}.
\newblock
\showISBNx{9798400700507}
\href{https://doi.org/10.1145/3576915.3624401}{doi:\nolinkurl{10.1145/3576915.3624401}}


\bibitem[Guo et~al\mbox{.}(2022)]%
        {guo2022unixcoder}
\bibfield{author}{\bibinfo{person}{Daya Guo}, \bibinfo{person}{Shuai Lu}, \bibinfo{person}{Nan Duan}, \bibinfo{person}{Yanlin Wang}, \bibinfo{person}{Ming Zhou}, {and} \bibinfo{person}{Jian Yin}.} \bibinfo{year}{2022}\natexlab{}.
\newblock \showarticletitle{Unixcoder: Unified cross-modal pre-training for code representation}.
\newblock \bibinfo{journal}{\emph{arXiv preprint arXiv:2203.03850}} (\bibinfo{year}{2022}).
\newblock


\bibitem[Guo et~al\mbox{.}(2025)]%
        {guo2025deepseek}
\bibfield{author}{\bibinfo{person}{Daya Guo}, \bibinfo{person}{Dejian Yang}, \bibinfo{person}{Haowei Zhang}, \bibinfo{person}{Junxiao Song}, \bibinfo{person}{Ruoyu Zhang}, \bibinfo{person}{Runxin Xu}, \bibinfo{person}{Qihao Zhu}, \bibinfo{person}{Shirong Ma}, \bibinfo{person}{Peiyi Wang}, \bibinfo{person}{Xiao Bi}, {et~al\mbox{.}}} \bibinfo{year}{2025}\natexlab{}.
\newblock \showarticletitle{Deepseek-r1: Incentivizing reasoning capability in llms via reinforcement learning}.
\newblock \bibinfo{journal}{\emph{arXiv preprint arXiv:2501.12948}} (\bibinfo{year}{2025}).
\newblock


\bibitem[Guo et~al\mbox{.}(2024)]%
        {guo2024deepseek}
\bibfield{author}{\bibinfo{person}{Daya Guo}, \bibinfo{person}{Qihao Zhu}, \bibinfo{person}{Dejian Yang}, \bibinfo{person}{Zhenda Xie}, \bibinfo{person}{Kai Dong}, \bibinfo{person}{Wentao Zhang}, \bibinfo{person}{Guanting Chen}, \bibinfo{person}{Xiao Bi}, \bibinfo{person}{Yu Wu}, \bibinfo{person}{YK Li}, {et~al\mbox{.}}} \bibinfo{year}{2024}\natexlab{}.
\newblock \showarticletitle{DeepSeek-Coder: When the Large Language Model Meets Programming--The Rise of Code Intelligence}.
\newblock \bibinfo{journal}{\emph{arXiv preprint arXiv:2401.14196}} (\bibinfo{year}{2024}).
\newblock


\bibitem[Hu et~al\mbox{.}(2024)]%
        {hu2024empirical}
\bibfield{author}{\bibinfo{person}{Jinchang Hu}, \bibinfo{person}{Lyuye Zhang}, \bibinfo{person}{Chengwei Liu}, \bibinfo{person}{Sen Yang}, \bibinfo{person}{Song Huang}, {and} \bibinfo{person}{Yang Liu}.} \bibinfo{year}{2024}\natexlab{}.
\newblock \showarticletitle{Empirical Analysis of Vulnerabilities Life Cycle in Golang Ecosystem}. In \bibinfo{booktitle}{\emph{Proceedings of the IEEE/ACM 46th International Conference on Software Engineering}}. \bibinfo{pages}{1--13}.
\newblock


\bibitem[Izacard and Grave(2021)]%
        {izacard-grave-2021-leveraging}
\bibfield{author}{\bibinfo{person}{Gautier Izacard} {and} \bibinfo{person}{Edouard Grave}.} \bibinfo{year}{2021}\natexlab{}.
\newblock \showarticletitle{Leveraging Passage Retrieval with Generative Models for Open Domain Question Answering}. In \bibinfo{booktitle}{\emph{Proceedings of the 16th Conference of the European Chapter of the Association for Computational Linguistics: Main Volume}}, \bibfield{editor}{\bibinfo{person}{Paola Merlo}, \bibinfo{person}{Jorg Tiedemann}, {and} \bibinfo{person}{Reut Tsarfaty}} (Eds.). \bibinfo{publisher}{Association for Computational Linguistics}, \bibinfo{address}{Online}, \bibinfo{pages}{874--880}.
\newblock
\href{https://doi.org/10.18653/v1/2021.eacl-main.74}{doi:\nolinkurl{10.18653/v1/2021.eacl-main.74}}


\bibitem[Kim et~al\mbox{.}(2019)]%
        {kim2019rvfuzzer}
\bibfield{author}{\bibinfo{person}{Taegyu Kim}, \bibinfo{person}{Chung~Hwan Kim}, \bibinfo{person}{Junghwan Rhee}, \bibinfo{person}{Fan Fei}, \bibinfo{person}{Zhan Tu}, \bibinfo{person}{Gregory Walkup}, \bibinfo{person}{Xiangyu Zhang}, \bibinfo{person}{Xinyan Deng}, {and} \bibinfo{person}{Dongyan Xu}.} \bibinfo{year}{2019}\natexlab{}.
\newblock \showarticletitle{$\{$RVFuzzer$\}$: Finding input validation bugs in robotic vehicles through $\{$Control-Guided$\}$ testing}. In \bibinfo{booktitle}{\emph{28th USENIX Security Symposium (USENIX Security 19)}}. \bibinfo{pages}{425--442}.
\newblock


\bibitem[Kruse et~al\mbox{.}(2022)]%
        {kruse2022multi}
\bibfield{author}{\bibinfo{person}{Rudolf Kruse}, \bibinfo{person}{Sanaz Mostaghim}, \bibinfo{person}{Christian Borgelt}, \bibinfo{person}{Christian Braune}, {and} \bibinfo{person}{Matthias Steinbrecher}.} \bibinfo{year}{2022}\natexlab{}.
\newblock \showarticletitle{Multi-layer perceptrons}.
\newblock In \bibinfo{booktitle}{\emph{Computational intelligence: a methodological introduction}}. \bibinfo{publisher}{Springer}, \bibinfo{pages}{53--124}.
\newblock


\bibitem[Li et~al\mbox{.}(2021)]%
        {li2021vuldeelocator}
\bibfield{author}{\bibinfo{person}{Zhen Li}, \bibinfo{person}{Deqing Zou}, \bibinfo{person}{Shouhuai Xu}, \bibinfo{person}{Zhaoxuan Chen}, \bibinfo{person}{Yawei Zhu}, {and} \bibinfo{person}{Hai Jin}.} \bibinfo{year}{2021}\natexlab{}.
\newblock \showarticletitle{Vuldeelocator: a deep learning-based fine-grained vulnerability detector}.
\newblock \bibinfo{journal}{\emph{IEEE Transactions on Dependable and Secure Computing}} \bibinfo{volume}{19}, \bibinfo{number}{4} (\bibinfo{year}{2021}), \bibinfo{pages}{2821--2837}.
\newblock


\bibitem[Meng et~al\mbox{.}(2018)]%
        {meng2018secure}
\bibfield{author}{\bibinfo{person}{Na Meng}, \bibinfo{person}{Stefan Nagy}, \bibinfo{person}{Danfeng Yao}, \bibinfo{person}{Wenjie Zhuang}, {and} \bibinfo{person}{Gustavo~Arango Argoty}.} \bibinfo{year}{2018}\natexlab{}.
\newblock \showarticletitle{Secure coding practices in java: Challenges and vulnerabilities}. In \bibinfo{booktitle}{\emph{Proceedings of the 40th International Conference on Software Engineering}}.
\newblock


\bibitem[{National Institute of Standards and Technology}(2024)]%
        {nist_csrc_vulnerability}
\bibfield{author}{\bibinfo{person}{{National Institute of Standards and Technology}}.} \bibinfo{year}{2024}\natexlab{}.
\newblock \bibinfo{title}{Software Vulnerability}.
\newblock
\urldef\tempurl%
\url{https://csrc.nist.gov/glossary/term/software_vulnerability}
\showURL{%
\tempurl}


\bibitem[of~Standards and Technology(2024)]%
        {nvd}
\bibfield{author}{\bibinfo{person}{National~Institute of Standards} {and} \bibinfo{person}{Technology}.} \bibinfo{year}{2024}\natexlab{}.
\newblock \bibinfo{title}{{National Vulnerability Database (NVD)}}.
\newblock
\urldef\tempurl%
\url{https://nvd.nist.gov/}
\showURL{%
\tempurl}


\bibitem[OpenAI(2022)]%
        {chatgpt2022}
\bibfield{author}{\bibinfo{person}{OpenAI}.} \bibinfo{year}{2022}\natexlab{}.
\newblock \bibinfo{title}{ChatGPT: Optimizing Language Models for Dialogue.}
\newblock
\urldef\tempurl%
\url{https://openai.com/blog/chatgpt}
\showURL{%
\tempurl}


\bibitem[OpenAI(2024a)]%
        {openai2024}
\bibfield{author}{\bibinfo{person}{OpenAI}.} \bibinfo{year}{2024}\natexlab{a}.
\newblock
\urldef\tempurl%
\url{https://openai.com/}
\showURL{%
\tempurl}


\bibitem[OpenAI(2024b)]%
        {openai2024gpt4o}
\bibfield{author}{\bibinfo{person}{OpenAI}.} \bibinfo{year}{2024}\natexlab{b}.
\newblock \bibinfo{title}{GPT-4o: A Flagship Model by OpenAI}.
\newblock
\urldef\tempurl%
\url{https://openai.com/index/gpt-4o-and-more-tools-to-chatgpt-free}
\showURL{%
\tempurl}


\bibitem[OpenAI(2024c)]%
        {openai_o1_2024}
\bibfield{author}{\bibinfo{person}{OpenAI}.} \bibinfo{year}{2024}\natexlab{c}.
\newblock \bibinfo{title}{OpenAI o1}.
\newblock
\urldef\tempurl%
\url{https://openai.com/o1/}
\showURL{%
\tempurl}


\bibitem[Purba et~al\mbox{.}(2023)]%
        {purba2023software}
\bibfield{author}{\bibinfo{person}{Moumita~Das Purba}, \bibinfo{person}{Arpita Ghosh}, \bibinfo{person}{Benjamin~J Radford}, {and} \bibinfo{person}{Bill Chu}.} \bibinfo{year}{2023}\natexlab{}.
\newblock \showarticletitle{Software vulnerability detection using large language models}. In \bibinfo{booktitle}{\emph{2023 IEEE 34th International Symposium on Software Reliability Engineering Workshops (ISSREW)}}. IEEE, \bibinfo{pages}{112--119}.
\newblock


\bibitem[Roziere et~al\mbox{.}(2023)]%
        {roziere2023code}
\bibfield{author}{\bibinfo{person}{Baptiste Roziere}, \bibinfo{person}{Jonas Gehring}, \bibinfo{person}{Fabian Gloeckle}, \bibinfo{person}{Sten Sootla}, \bibinfo{person}{Itai Gat}, \bibinfo{person}{Xiaoqing~Ellen Tan}, \bibinfo{person}{Yossi Adi}, \bibinfo{person}{Jingyu Liu}, \bibinfo{person}{Tal Remez}, \bibinfo{person}{J{\'e}r{\'e}my Rapin}, {et~al\mbox{.}}} \bibinfo{year}{2023}\natexlab{}.
\newblock \showarticletitle{Code llama: Open foundation models for code}.
\newblock \bibinfo{journal}{\emph{arXiv preprint arXiv:2308.12950}} (\bibinfo{year}{2023}).
\newblock


\bibitem[Shiri~Harzevili et~al\mbox{.}(2024)]%
        {shiri2024systematic}
\bibfield{author}{\bibinfo{person}{Nima Shiri~Harzevili}, \bibinfo{person}{Alvine Boaye~Belle}, \bibinfo{person}{Junjie Wang}, \bibinfo{person}{Song Wang}, \bibinfo{person}{Zhen~Ming Jiang}, {and} \bibinfo{person}{Nachiappan Nagappan}.} \bibinfo{year}{2024}\natexlab{}.
\newblock \showarticletitle{A systematic literature review on automated software vulnerability detection using machine learning}.
\newblock \bibinfo{journal}{\emph{Comput. Surveys}} \bibinfo{volume}{57}, \bibinfo{number}{3} (\bibinfo{year}{2024}), \bibinfo{pages}{1--36}.
\newblock


\bibitem[Tian et~al\mbox{.}(2024)]%
        {tian2024large}
\bibfield{author}{\bibinfo{person}{Zhao Tian}, \bibinfo{person}{Honglin Shu}, \bibinfo{person}{Dong Wang}, \bibinfo{person}{Xuejie Cao}, \bibinfo{person}{Yasutaka Kamei}, {and} \bibinfo{person}{Junjie Chen}.} \bibinfo{year}{2024}\natexlab{}.
\newblock \showarticletitle{Large Language Models for Equivalent Mutant Detection: How Far Are We?}. In \bibinfo{booktitle}{\emph{Proceedings of the 33rd ACM SIGSOFT International Symposium on Software Testing and Analysis}}. \bibinfo{pages}{1733--1745}.
\newblock


\bibitem[Touvron et~al\mbox{.}(2023)]%
        {touvron2023llama}
\bibfield{author}{\bibinfo{person}{Hugo Touvron}, \bibinfo{person}{Louis Martin}, \bibinfo{person}{Kevin Stone}, \bibinfo{person}{Peter Albert}, \bibinfo{person}{Amjad Almahairi}, \bibinfo{person}{Yasmine Babaei}, \bibinfo{person}{Nikolay Bashlykov}, \bibinfo{person}{Soumya Batra}, \bibinfo{person}{Prajjwal Bhargava}, \bibinfo{person}{Shruti Bhosale}, {et~al\mbox{.}}} \bibinfo{year}{2023}\natexlab{}.
\newblock \showarticletitle{Llama 2: Open foundation and fine-tuned chat models}.
\newblock \bibinfo{journal}{\emph{arXiv preprint arXiv:2307.09288}} (\bibinfo{year}{2023}).
\newblock


\bibitem[Wang et~al\mbox{.}(2023b)]%
        {wang2023reef}
\bibfield{author}{\bibinfo{person}{Chaozheng Wang}, \bibinfo{person}{Zongjie Li}, \bibinfo{person}{Yun Pena}, \bibinfo{person}{Shuzheng Gao}, \bibinfo{person}{Sirong Chen}, \bibinfo{person}{Shuai Wang}, \bibinfo{person}{Cuiyun Gao}, {and} \bibinfo{person}{Michael~R Lyu}.} \bibinfo{year}{2023}\natexlab{b}.
\newblock \showarticletitle{Reef: A framework for collecting real-world vulnerabilities and fixes}. In \bibinfo{booktitle}{\emph{2023 38th IEEE/ACM International Conference on Automated Software Engineering (ASE)}}. IEEE, \bibinfo{pages}{1952--1962}.
\newblock


\bibitem[Wang et~al\mbox{.}(2024)]%
        {wang2024software}
\bibfield{author}{\bibinfo{person}{Junjie Wang}, \bibinfo{person}{Yuchao Huang}, \bibinfo{person}{Chunyang Chen}, \bibinfo{person}{Zhe Liu}, \bibinfo{person}{Song Wang}, {and} \bibinfo{person}{Qing Wang}.} \bibinfo{year}{2024}\natexlab{}.
\newblock \showarticletitle{Software testing with large language models: Survey, landscape, and vision}.
\newblock \bibinfo{journal}{\emph{IEEE Transactions on Software Engineering}} (\bibinfo{year}{2024}).
\newblock


\bibitem[Wang et~al\mbox{.}(2023a)]%
        {wang2023codet5+}
\bibfield{author}{\bibinfo{person}{Yue Wang}, \bibinfo{person}{Hung Le}, \bibinfo{person}{Akhilesh~Deepak Gotmare}, \bibinfo{person}{Nghi~DQ Bui}, \bibinfo{person}{Junnan Li}, {and} \bibinfo{person}{Steven~CH Hoi}.} \bibinfo{year}{2023}\natexlab{a}.
\newblock \showarticletitle{Codet5+: Open code large language models for code understanding and generation}.
\newblock \bibinfo{journal}{\emph{arXiv preprint arXiv:2305.07922}} (\bibinfo{year}{2023}).
\newblock


\bibitem[Wang et~al\mbox{.}(2021)]%
        {wang2021codet5}
\bibfield{author}{\bibinfo{person}{Yue Wang}, \bibinfo{person}{Weishi Wang}, \bibinfo{person}{Shafiq Joty}, {and} \bibinfo{person}{Steven~CH Hoi}.} \bibinfo{year}{2021}\natexlab{}.
\newblock \showarticletitle{Codet5: Identifier-aware unified pre-trained encoder-decoder models for code understanding and generation}.
\newblock \bibinfo{journal}{\emph{arXiv preprint arXiv:2109.00859}} (\bibinfo{year}{2021}).
\newblock


\bibitem[{WhiteSource}(2022)]%
        {whitesource2022mend}
\bibfield{author}{\bibinfo{person}{{WhiteSource}}.} \bibinfo{year}{2022}\natexlab{}.
\newblock \bibinfo{title}{Mend bolt}.
\newblock
\urldef\tempurl%
\url{https://www.mend.io/free-developer-tools/bolt}
\showURL{%
\tempurl}


\bibitem[Wolf et~al\mbox{.}(2019)]%
        {wolf2019huggingface}
\bibfield{author}{\bibinfo{person}{Thomas Wolf}, \bibinfo{person}{Lysandre Debut}, \bibinfo{person}{Victor Sanh}, \bibinfo{person}{Julien Chaumond}, \bibinfo{person}{Clement Delangue}, \bibinfo{person}{Anthony Moi}, \bibinfo{person}{Pierric Cistac}, \bibinfo{person}{Tim Rault}, \bibinfo{person}{R{\'e}mi Louf}, \bibinfo{person}{Morgan Funtowicz}, {et~al\mbox{.}}} \bibinfo{year}{2019}\natexlab{}.
\newblock \showarticletitle{Huggingface's transformers: State-of-the-art natural language processing}.
\newblock \bibinfo{journal}{\emph{arXiv preprint arXiv:1910.03771}} (\bibinfo{year}{2019}).
\newblock


\bibitem[Wu et~al\mbox{.}(2022)]%
        {wu2022vulcnn}
\bibfield{author}{\bibinfo{person}{Yueming Wu}, \bibinfo{person}{Deqing Zou}, \bibinfo{person}{Shihan Dou}, \bibinfo{person}{Wei Yang}, \bibinfo{person}{Duo Xu}, {and} \bibinfo{person}{Hai Jin}.} \bibinfo{year}{2022}\natexlab{}.
\newblock \showarticletitle{Vulcnn: An image-inspired scalable vulnerability detection system}. In \bibinfo{booktitle}{\emph{Proceedings of the 44th International Conference on Software Engineering}}. \bibinfo{pages}{2365--2376}.
\newblock


\bibitem[Xu et~al\mbox{.}(2022)]%
        {xu2022insight}
\bibfield{author}{\bibinfo{person}{Meiqiu Xu}, \bibinfo{person}{Ying Wang}, \bibinfo{person}{Shing-Chi Cheung}, \bibinfo{person}{Hai Yu}, {and} \bibinfo{person}{Zhiliang Zhu}.} \bibinfo{year}{2022}\natexlab{}.
\newblock \showarticletitle{Insight: Exploring cross-ecosystem vulnerability impacts}. In \bibinfo{booktitle}{\emph{Proceedings of the 37th IEEE/ACM International Conference on Automated Software Engineering}}. \bibinfo{pages}{1--13}.
\newblock


\bibitem[Yang et~al\mbox{.}(2024)]%
        {yang2024qwen2}
\bibfield{author}{\bibinfo{person}{An Yang}, \bibinfo{person}{Baosong Yang}, \bibinfo{person}{Beichen Zhang}, \bibinfo{person}{Binyuan Hui}, \bibinfo{person}{Bo Zheng}, \bibinfo{person}{Bowen Yu}, \bibinfo{person}{Chengyuan Li}, \bibinfo{person}{Dayiheng Liu}, \bibinfo{person}{Fei Huang}, \bibinfo{person}{Haoran Wei}, {et~al\mbox{.}}} \bibinfo{year}{2024}\natexlab{}.
\newblock \showarticletitle{Qwen2. 5 technical report}.
\newblock \bibinfo{journal}{\emph{arXiv preprint arXiv:2412.15115}} (\bibinfo{year}{2024}).
\newblock


\bibitem[Yuan et~al\mbox{.}(2023)]%
        {yuan2023enhancing}
\bibfield{author}{\bibinfo{person}{Bin Yuan}, \bibinfo{person}{Yifan Lu}, \bibinfo{person}{Yilin Fang}, \bibinfo{person}{Yueming Wu}, \bibinfo{person}{Deqing Zou}, \bibinfo{person}{Zhen Li}, \bibinfo{person}{Zhi Li}, {and} \bibinfo{person}{Hai Jin}.} \bibinfo{year}{2023}\natexlab{}.
\newblock \showarticletitle{Enhancing deep learning-based vulnerability detection by building behavior graph model}. In \bibinfo{booktitle}{\emph{2023 IEEE/ACM 45th International Conference on Software Engineering (ICSE)}}. IEEE, \bibinfo{pages}{2262--2274}.
\newblock


\bibitem[Zerouali et~al\mbox{.}(2022)]%
        {zerouali2022impact}
\bibfield{author}{\bibinfo{person}{Ahmed Zerouali}, \bibinfo{person}{Tom Mens}, \bibinfo{person}{Alexandre Decan}, {and} \bibinfo{person}{Coen De~Roover}.} \bibinfo{year}{2022}\natexlab{}.
\newblock \showarticletitle{On the impact of security vulnerabilities in the npm and RubyGems dependency networks}.
\newblock \bibinfo{journal}{\emph{Empirical Software Engineering}} \bibinfo{volume}{27}, \bibinfo{number}{5} (\bibinfo{year}{2022}), \bibinfo{pages}{107}.
\newblock


\end{thebibliography}
